%% file: main.tex
\documentclass[conference,letterpaper]{IEEEtran}

\usepackage[utf8]{inputenc} 
\usepackage[T1]{fontenc}
\usepackage{url}
\usepackage{ifthen}
\usepackage{cite}
\usepackage{mathtools}
\usepackage{bm,hyperref}
\usepackage{xcolor}
\usepackage{amsmath} 
\usepackage{amssymb}
\begin{document}
\title{An Information-Theoretic Efficient Capacity Region for Multi-User Interference Channel} 

\author{%
  \IEEEauthorblockN{Sagnik Bhattacharya*, Abhiram R. Gorle*, Muhammad Ahmed Mohsin, John M. Cioffi}
  \IEEEauthorblockA{Department of Electrical Engineering, Stanford University, USA \\
                    Emails: \{sagnikb, abhiramg, muahmed, cioffi\}@stanford.edu \\
                    * Equal contribution}
}

\maketitle

\begin{abstract} 

We investigate the capacity region of multi-user interference channels (IC), where each user encodes multiple sub-user components. By unifying chain-rule decomposition with the Entropy Power Inequality (EPI), we reason that single-user Gaussian codebooks suffice to achieve optimal performance, thus obviating any need for intricate auxiliary variables or joint typicality arguments. Our partial-MAC formulation enumerates sub-user decoding orders while only imposing constraints for sub-users actually decoded. This significantly reduces complexity relative to enumerating all subsets or bruteforcing over all successive interference cancellation (SIC) decoding order combinations at all receivers. This leads to a finite but comprehensive construction of all achievable rate tuples under sum-power constraints, while guaranteeing that each receiver fully recovers its intended sub-user signals. Consequently, known single-user Gaussian capacity results generalize naturally to multi-user scenarios, revealing a cohesive framework for analyzing multi-user IC. Our results thus offer a streamlined, tractable pathway for designing next-generation cell-free wireless networks that rely on IC mechanisms, efficiently exploiting interference structure while minimizing overhead. Overall, this provides a unifying perspective.
\end{abstract}

\input{intro}

\input{multi-user-ic}

\input {epi}

\input {ic-capacity}

\input {reformulated-efficient}

\input {results}

\input {conclusion}


\bibliography{main}
\bibliographystyle{IEEEtran}

\newpage

\appendix
\input{appendix}
\label{appendix:refinedEPI}



\end{document}

%% file: intro.tex
\section{Introduction}
\label{sec:intro}

In a multi-user wireless network, an interference channel (IC) consists of \(U\) independent transmitters and \(U\) independent receivers. Each transmitter encodes \(U\) sub-user components, resulting in \(U^2\) sub-user signals across the channel. Each receiver employs successive interference cancellation (SIC) to decode its intended sub-user components and, optionally, some interfering sub-user components, treating the rest as noise. Unlike traditional BC or MAC scenarios where users or receivers are grouped under a single base station (BS) or access point (AP), the IC lacks such coordination, rendering the analysis considerably more complex.

The IC model is especially relevant in modern applications such as distributed AI-enabled augmented/virtual reality (AR/VR), federated learning, and edge training, all of which rely on next-generation wireless networks. Consequently, the achievable rate-capacity region and power-efficient transmission strategies are paramount. Consequently, characterizing the achievable rate-capacity region and developing power minimization strategies are key challenges, first studied in \cite{HK,Sato81} (this work subsumes the famous Han-Kobayashi region for a two-user interference-channel scheme). Our approach generalizes to \(U\)-user ICs by introducing up to \(U\) sub-streams per user, enumerating partial decoding orders (partial-MAC principle), and showing that strictly single-user Gaussian codebooks suffice to characterize the region. This eliminates the need for sophisticated auxiliary variables and simultaneously accommodates an arbitrary number of users.

While the capacity regions for multiple-access (MAC) and broadcast (BC) channels are well-established with Gaussian inputs being optimal under power constraints~\cite{shannon1948mathematical}, the capacity region for general multi-user interference channels (ICs) remains unresolved~\cite{open-problem1, open-problem2, open-problem3}. A primary challenge lies in determining whether Gaussian distributions are optimal for all sub-user components within ICs~\cite{mac-bc-gaussian}. In this work, we leverage a reformulated entropy power inequality (EPI) \cite{epi} to show that in additive Gaussian noise channels, \emph{each} sub-user component is indeed optimally Gaussian.

This result leads to a comprehensive mathematical characterization of the multi-user interference channel (IC) achievable rate-capacity region under a sum-power constraint. Importantly, for additive Gaussian noise channels, the multi-user capacity region herein covers or surpasses previously known achievable regions. This is achieved through a finite but comprehensive construction that unifies the chain rule with entropy-power arguments, providing an explicit and direct characterization of all achievable rate tuples. Thus, the proposed approach subsumes existing inner bounds for AWGN interference channels. The capacity region under this constraint represents all achievable data rate combinations across users, ensuring reliable decoding at respective receivers while satisfying a total power budget. We show that the capacity region for the multi-user IC is \textit{convex} and provide a detailed formulation for constructing this convex capacity region.

Now, with this capacity region characterized, there remains the challenge of minimizing the total power consumption for a multi-user IC while satisfying each user's minimum data-rate requirements for each user. This optimization problem is inherently difficult because it involves exploring all possible receiver decoding-order combinations, resulting in a combinatorial search space of $(U^2!)^U$. To address this computational challenge, a novel algorithm, \texttt{minPIC} decouples the capacity region characterization from the decoding order selection. This decoupling leads to significant runtime improvements, achieving an order-of-magnitude faster convergence compared to a brute-force approach. The simulation results demonstrate the capacity region construction for various IC examples and showcase the efficacy of the proposed sum-power minimization solutions using the \texttt{minPIC} algorithm.



We propose a dynamic time-sharing algorithm that varies the decoding orders across receivers over time, achieving lower average power consumption than static decoding orders while fulfilling data rate requirements. Simulations compare both \texttt{minPIC} and time-sharing schemes to orthogonal multiple access (OMA) for a $2\times 2$ IC, currently standard in cellular \cite{5g} and Wi-Fi \cite{wifi} systems. Results for an 80 MHz Wi-Fi channel demonstrate that the proposed approach achieves higher data rates for a given power budget or, equivalently, satisfies data rate requirements with lower total power consumption compared to OMA-based solutions \cite{optimal-power-subcarrier-mac, fwa}. These findings highlight the superiority of the proposed methodology over existing standards.

%% file: multi-user-ic.tex
\section{Multi-user Interference Channel}
\label{sec:multi-user-ic}
As stated earlier, the multi-user interference channel (IC) has a setup where \(U\) users transmit information to \(U\) receivers simultaneously over a shared medium. Unlike multiple-access channels (MACs) and broadcast channels (BCs), in interference channels, transmitters and receivers cannot co-process transmit nor receive signals, leading to inter-user interference that complicates the communication process. Thus, the interference channel evolves into a more universal and challenging communication framework, encapsulating many essential real-world wireless system elements.

The capacity region is the set of all reliably achievable rate tuples $(R_1, R_2, \dots, R_U)$, where $R_k$ represents the $k^{\text{th}}$ user's total data rate, subject to the decoding constraints and sum-power constraints.

For a system with \(U\) users and \(U\) receivers, the received signal at the $i^{\text{th}}$ receiver is:
\begin{equation}
    y_i = \sum_{k=1}^U \sum_{j=1}^U h_{i,k} x_{k,j} + z_i,
\end{equation}
where $h_{i,k}$ denotes the channel gain from the $k^{\text{th}}$ transmitter to the $i^{\text{th}}$ receiver, $x_{k,j}$ represents the $j^{\text{th}}$ sub-user component of the $k^{\text{th}}$ user, and $z_i \sim \mathcal{N}(0, \sigma_i^2)$ is the additive Gaussian noise.

\subsection{Sub-User Components and Decoding Orders}

\begin{figure}
    \centering
    \includegraphics[width=\linewidth]{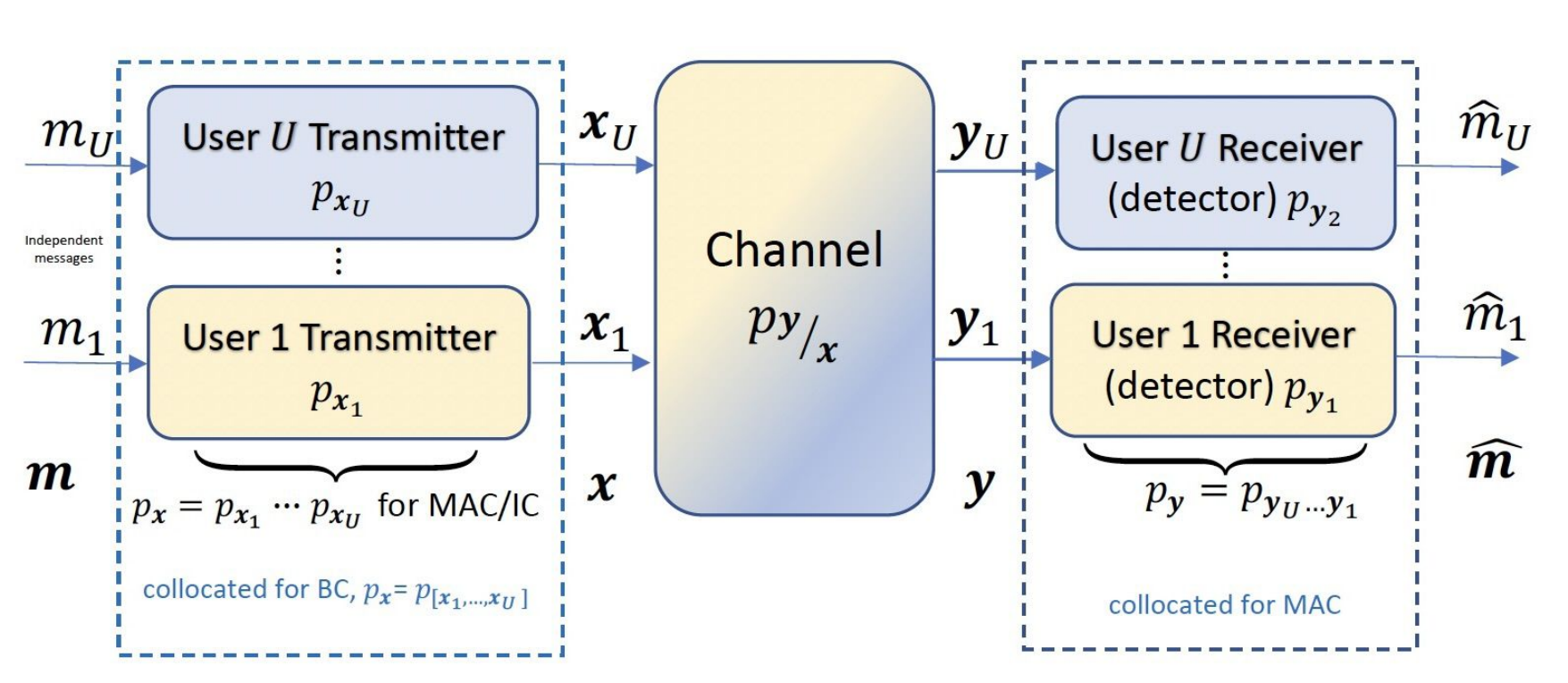}
    \caption{A 2-user interference channel}
    \label{fig:ic}
\end{figure}

Each IC user generates \(U\) independently coded sub-user components, leading to a total of \(U^2\) sub-user components transmitted across the channel. At each receiver, successive interference cancellation (SIC) decodes the received signals. The \(i^{\text{th}}\) receiver must decode all \(U\) sub-user components transmitted by the \(i^{\text{th}}\) user. Additionally, the receiver decodes a subset of the sub-user components transmitted by other users that appear in the SIC decoding order before the \(i^{\text{th}}\) user's components. Any sub-user components from other users that appear later in the decoding order are treated as noise. This intricate decoding structure highlights the key challenge of designing optimal decoding orders and power allocations to maximize data rates and minimize interference. Fig.~\ref{fig:ic} outlines a typical interference channel scenario.

Notably, enumerating all possible global SIC orders across \( U \) receivers requires considering \( (U^2!)^U \) sequences. Each receiver’s local SIC order among the \( U^2 \) sub-user components determines which sub-users are decoded or treated as noise. This finite, albeit potentially large, design eliminates the need for continuous auxiliary variables or joint-typical-set arguments, and is a natural generalization of prior “common/private” strategies \cite{HK}.

\subsection{Comparison with MAC and BC Channels}
The interference channel generalizes both the MAC and BC channels, which can be viewed as degenerate cases of the IC:
\begin{itemize}
    \item In a MAC, multiple users transmit to a single receiver. This effectively groups the receivers into one entity, reducing the total number of independently decoded components from \(U^2\) to \(U\). This simplification significantly reduces decoding complexity, as the total number of decoding orders decreases from \((U^2!)^U\) to \(U!\).
    \item In a BC, a single transmitter communicates with multiple receivers. Here, the transmitters group together, similarly reducing complexity, as in the MAC case.
\end{itemize}

For MACs and BCs \cite{Yu2004}, \cite{dual2004}, it is well-established that assuming Gaussian input distributions for all sub-user components is optimal because Gaussian distributions maximize entropy for a given power constraint. This greatly facilitates these channels' capacity region characterization.

\subsection{Challenges in IC Capacity Region Characterization}
In contrast to MACs and BCs, the general IC scenario is significantly more complex because of the lack of processing among users and receivers. A major roadblock in mathematically characterizing the IC capacity region is determination of the optimal input distributions.
The optimality of Gaussian signaling in Multiple Access Channels (MAC) and Broadcast Channels (BC) under AWGN conditions is well-established in \cite{CoverThomas2006, Cover1972, Slepian1973, Shannon1948}, leveraging the maximum differential entropy property of Gaussian distributions to maximize mutual information. Unlike the MAC and BC, where Gaussian inputs are provably optimal, it is not immediately evident whether Gaussian signaling achieves the capacity region for the IC. 

In this work, we make direct use of the chain rule for mutual information, along with Shannon’s entropy-power inequality (EPI), to avoid any need for introducing continuous 
auxiliary random variables or complicated joint typical-set arguments. This ensures a finite (though possibly large) specification of all achievable points, thereby offering an explicit construction that matches or exceeds previously known bounds.


\subsection{Power Minimization Under Rate Constraints}
Beyond capacity region characterization, another critical challenge is minimizing the total power consumption while satisfying minimum data rate requirements for all users in the IC. Despite its importance, there has been limited prior address of this problem for multi-user ICs. Most existing approaches focus on simpler scenarios, under simplifying assumptions on user coordination or location, etc., leaving a significant gap in practical algorithms for power-efficient IC communication. The following section proceeds to prove the Gaussian distribution's optimality on all IC sub-user components using the entropy power inequality (Costa, \cite{epi}).




%% file: epi.tex
\section{Gaussian assumption on sub-user components}
\label{sec:epi}

The Entropy Power Inequality (EPI) is a fundamental result introduced by Shannon in \cite{shannon1948mathematical}, that provides a relationship between the differential entropy of independent random variables and the entropy of their sum. For continuous random variables \(X\) and \(Y\) with finite differential entropies \(h(X)\) and \(h(Y)\), the primitive version of the EPI states:

\begin{equation}
    N(X + Y) \geq N(X) + N(Y),
\end{equation}
where \(N(X) \coloneq \frac{1}{2\pi e} e^{2h(X)}\) is the entropy power of \(X\). The equality holds if and only if \(X\) and \(Y\) are Gaussian with fixed, known covariance matrices.

Shannon's original formulation holds strictly for continuous random variables in Euclidean spaces and assumes that the sum \(X+Y\) exists in the same space. The classical EPI, rooted in Shannon's work on Gaussian channels and additive noise, has since been extended to handle dependencies, non-Gaussian distributions, and refine entropy bounds for broader scenarios in \cite{epi}, \cite{Stam1959}, \cite{Blachman1965}, \cite{Dembo1991}, \cite{feder1993}, \cite{verdu2006}.

\subsection{Strengthened Versions of the EPI}
Modern refinements of the EPI have possibly extended its application to non-Gaussian settings and its relationship to mutual information and estimation theory. More notably, the strengthened EPI incorporates auxiliary terms that improve the bound, particularly when \(X\) and \(Y\) deviate from the Gaussian setting \cite{strengtheningEPI}. Elegant relations between the MMSE and the derivative of the mutual information with the signal-to-noise ratio (SNR) have been explored, yielding alternative proofs and broader implications for Gaussian channels in \cite{verdu2006}, \cite{2004verdguo}. The reformulated EPI plays a critical role in the analysis of interference channels, particularly in justifying the optimality of Gaussian input distributions.

\subsection{Proof of the EPI: A Differential Entropy Approach}
Unlike previous EPI proofs (presented in \cite{Stam1959}, \cite{Dembo1991}), this proof does not hinge on Fisher’s information. One of the most insightful approaches to proving the EPI leverages the relationship between mutual information and the minimum mean-square error (MMSE) outlined in \cite{2004verdguo}. For a scalar Gaussian channel with input \(X\) and output \(Y = \sqrt{\text{$\gamma$}} \, X + N\), where \(N \sim \mathcal{N}(0, 1)\), the derivative of the mutual information \(I(X; Y)\) with respect to the signal-to-noise ratio (in this case, $\gamma$) is proportional to the MMSE:

\begin{equation}
    \frac{\partial I(X; Y)}{\partial \text{$\gamma$}} = \frac{1}{2} \mathrm{mmse}(\text{$\gamma$}).
\end{equation}

By integrating this relationship and applying the chain rule for differential entropy, the EPI emerges naturally as a consequence of the Gaussian distribution minimizing the MMSE at a fixed variance. \cite{2004verdguo}

\subsection{Proof of the EPI: Forney’s Crypto Lemma and Mixing}
Another elegant proof involves Forney’s crypto lemma from \cite{forney2004}, which highlights the Gaussian distribution's role in maximizing entropy under linear transformations. When random variables are mixed repeatedly via linear operations and additive Gaussian noise, the resulting distribution converges towards a Gaussian due to the Central Limit Theorem and the entropy maximization property of Gaussians. This insight establishes that the entropy power of sums increases, formalizing the EPI.

\subsection{Implications for Gaussianity and Mixing}

The Entropy Power Inequality (EPI) implies that when independent, continuously distributed random variables are successively combined with Gaussian noise, their distribution increasingly resembles a Gaussian shape. This trend is rigorously supported by the Entropy Power Inequality (EPI) as stated in the above sections because the Gaussian distribution uniquely maximizes entropy among distributions with the same covariance, and each addition of noise progressively nudges the overall distribution closer to Gaussian. 

Moreover, this insight aligns with the Entropic Central Limit Theorem \cite{barron1986}, which refines the classical Central Limit Theorem by showing that the differential entropy of normalized sums converges to that of the limiting Gaussian. In essence, the Gaussian's high-entropy characteristic, combined with the EPI’s assertion of entropy growth under convolution, explains why sums of independent random variables—smoothed by Gaussian perturbations—approach Gaussian-like behavior.



Building on our setup from \ref{sec:multi-user-ic}, 
during successive interference cancellation (SIC), the decoding process involves iteratively decoding and subtracting certain sub-user components, leaving the remaining components as interference. 
\\
Let $\mathcal{D}_i^{(m)}$ represent the set of sub-user components decoded by the $i^{\text{th}}$ receiver in the $m^{\text{th}}$ step of SIC, and let $\mathcal{I}_i^{(m)}$ represent the interference from components not yet decoded. Then, the effective signal at each decoding step is:
\begin{equation}
    \tilde{y}_i^{(m)} = \sum_{(k,j) \in \mathcal{D}_i^{(m)}} h_{i,k} x_{k,j} + \sum_{(k',j') \in \mathcal{I}_i^{(m)}} h_{i,k'} x_{k',j'} + z_i.
\end{equation}

The EPI asserts that the entropy power of the mixture $\tilde{y}_i^{(m)}$ satisfies:
\begin{align}
    N(\tilde{y}_i^{(m)}) &\geq 
    \sum_{(k,j) \in \mathcal{D}_i^{(m)}}\! N(h_{i,k}x_{k,j}) 
    + \sum_{(k',j') \in \mathcal{I}_i^{(m)}}\! N(h_{i,k'}x_{k',j'}) 
    + N(z_i),
\end{align}

The equality holds if and only if all the constituent terms in the mixture are Gaussian.

\subsubsection{Optimality of Gaussian Inputs}

For Gaussian channels, maximizing the mutual information $I(x; y)$ between the input and output subject to a power constraint $P$ corresponds to maximizing the differential entropy $h(y)$. The EPI ensures that Gaussian inputs, which maximize entropy for a fixed variance, are optimal in this setting. Specifically, for any $x_{k,j}$ with power $p_{k,j}$, we have:
\begin{equation}
    h(x_{k,j}) \leq \frac{1}{2} \log(2\pi e p_{k,j}),
\end{equation}
with equality if and only if $x_{k,j} \sim \mathcal{N}(0, p_{k,j})$.

A particularly crucial consequence of the \textit{EPI-based chain-rule decomposition} is the following: any nonzero-variance additive Gaussian noise implies that optimal subuser signals are also Gaussian. The effective interference seen by each subuser includes the sum of leftover subusers treated as noise plus the actual AWGN term, which remains ‘\textit{sub-Gaussian}.’ Hence, a Gaussian-coded subuser maximizes mutual information under the power constraints. This shows that the Gaussian distribution is indeed optimal for all sub-user components in the Gaussian IC, generalizing well-known single-user capacity results to the multi-user setting.

\subsubsection{Reduction of Optimization Complexity}

Without the Gaussian assumption, deriving the capacity region would  typically involve optimizing over all possible input distributions $p_{x_{k,j}}$, a problem of prohibitive complexity due to the vast space of potential distributions. EPI invocation restricts the input distributions to Gaussian, significantly reducing the complexity. Mathematically, the achievable rate for sub-user component $(k,j)$ decoded at receiver $i$ is:
\begin{equation}
    R_{k,j}^{(i)} \leq \log\left(1 + \frac{|h_{i,k}|^2 p_{k,j}}{\sigma_i^2 + \sum_{(k',j') \in \mathcal{I}_i^{(m)}} |h_{i,k'}|^2 p_{k',j'}}\right),
\end{equation}
which assumes that both interference and desired signals are Gaussian.





%% file: ic-capacity.tex
\section{Interference Channel Capacity Region}
\label{sec:ic-capacity}
\subsection{Capacity Region Derivation}

Leveraging EPI validates the Gaussianity assumption for all sub-user components and permits the characterization of the capacity region of the multi-user interference channel (IC). 

\subsubsection{Intersection and Union Formulation}
The IC capacity region is the intersection of rate regions determined at each receiver. Each receiver forms a multiple-access channel (MAC) with the transmitters, and its achievable rate region is the union over all possible decoding orders of the sub-user components. Let $\mathcal{C}_i$ denote the achievable rate region at the $i^{\text{th}}$ receiver. The capacity region $\mathcal{C}_{\text{IC}}$ of the IC is given by:
\begin{equation}
\label{eq:partial-mac}
    \mathcal{C}_{\text{IC}} = \bigcap_{i=1}^U \mathcal{C}_i,
\end{equation}
where $\mathcal{C}_i$ is determined as:
\begin{equation}
    \mathcal{C}_i = \bigcup_{\pi_i} \left\{ (R_{k,j}) \; \bigg| \; \sum_{(k,j) \in \mathcal{D}_i} R_{k,j} \leq I(\mathcal{D}_i; y_i \mid \mathcal{I}_i), \; \forall \mathcal{D}_i \right\}.
\end{equation}

where,
\begin{itemize}
    \item $\pi_i$ is the decoding order at the $i^{\text{th}}$ receiver.
    \item $\mathcal{D}_i$ is the set of sub-user components decoded at the $i^{\text{th}}$ receiver.
    \item $\mathcal{I}_i$ represents the interference from sub-user components not decoded at the $i^{\text{th}}$ receiver.
\end{itemize}

The achievable rate for each sub-user component $(k,j)$ decoded at the $i^{\text{th}}$ receiver is:
\begin{equation}
    R_{k,j} \leq \log\left( 1 + \frac{|h_{i,k}|^2 p_{k,j}}{\sigma_i^2 + \sum_{(k',j') \in \mathcal{I}_i} |h_{i,k'}|^2 p_{k',j'}} \right),
\end{equation}
where $p_{k,j}$ is the power allocated to sub-user component $(k,j)$.

\subsection{Partial MAC Capacity Region}
While the above formulation in eq.~\eqref{eq:partial-mac} characterizes the IC capacity region as an intersection of MAC capacity regions across receivers, there is a critical distinction. For the $i^{\text{th}}$ IC receiver, only the sub-user components of the $i^{\text{th}}$ user must be fully decoded, whereas other sub-user components are decoded only if they fall within the decoding order required to decode the $i^{\text{th}}$ user's components. This creates a fundamental difference between the MAC capacity region for a single receiver and the MAC capacity regions solved as a part of the intersection for the IC region characterization. This modified MAC capacity region in Eq.~\eqref{eq:partial-mac} as the \emph{partial MAC} capacity region. 





%% file: reformulated-efficient.tex
\section{Efficient Power Minimization for Multi-user IC}
\label{sec:reformulated-efficient}
A design goal is to minimize the sum of the power allocated to all sub-user components across all users in a multi-user interference channel (IC), subject to each user's minimum rate requirements. This problem is fundamental in next-generation wireless applications (e.g., AR/VR, edge computing), where stringent data rates must be met while conserving energy.

From the capacity region characterization (Section~\ref{sec:ic-capacity}), the minimum sum-power problem can be formulated as
\begin{align}
\label{eq:opt}
    \min_{\{p_{k,j}\}} & \quad \sum_{k=1}^U \sum_{j=1}^U p_{k,j}, \\
    \text{s.t.} & \quad R_k \geq R_k^{\text{min}}, \quad \forall k \in \{1, 2, \dots, U\}, \\
    & \quad R_{k,j} \leq \log\left(1 + \frac{|h_{i,k}|^2 p_{k,j}}{\sigma_i^2 + \sum_{(k',j') \in \mathcal{I}_i} |h_{i,k'}|^2 p_{k',j'}} \right), \nonumber \\
    & \quad \text{and other constraints from } \mathcal{C}_{\text{IC}}.
\end{align}

where, $p_{k,j}$ is the power allocated to the $j^{\text{th}}$ sub-user component of the $k^{\text{th}}$ user, $R_k^{\text{min}}$ is the minimum required data rate for user $k$, and $\mathcal{C}_{\text{IC}}$ encodes the partial MAC constraints discussed earlier.

\subsection{Challenges in Decoding Order Combinations}
A major hurdle in the above formulation arises from the necessity to identify an SIC decoding order at each of the \(U\) receivers. Since each receiver observes \(U^2\) sub-user components, the total number of global decoding-order combinations scales as \(\bigl(U^2!\bigr)^U\), quickly becoming intractable for even moderate \(U\) (e.g., $U = 3$). A naive approach would require enumerating every possible permutation to find the minimum sum-power satisfying the user-rate constraints.


\subsection{Proposed Reformulation and Constraint Reduction}
To alleviate this bottleneck, we reformulate the partial MAC capacity region at each receiver by excluding constraints for sub-user subsets that are never decoded. This approach, referred to as “reformulated partial MAC,” removes explicit order dependence. Concretely, instead of enumerating \(\bigl(U^2!\bigr)^U\) decoding orders, we impose only those capacity constraints needed to ensure correctness for sub-users that \emph{must} be decoded. This yields a drastic reduction—from \(\bigl(U^2!\bigr)^U\) constraints to roughly \(U \bigl(2^{U^2} - 2^{\,U^2 - U}\bigr)\), all while preserving convexity in the sum-power allocation, therefore making the problem tractable for practical systems.

\subsection{Lagrangian-Based Optimization and SIC Decoupling}

While the decoding-order space is combinatorial ($(U^2!)^U$ possible orders), one can decouple the search for minimum sum-power from the explicit enumeration of these 
orders via a Lagrangian-based or “saddle-point” optimization. Concretely, one introduces dual variables for each user’s rate or power constraints, and updates the allocated sub-user powers in a convex-descent procedure. Only at the final stage, or via an intelligent search, does one identify the actual SIC orders used. If no single order meets the exact rate requirements, time-sharing among a small set of nearest “vertices” suffices. This approach dramatically reduces complexity in large-scale ICs.

In more rigorous terms, let \(\{\lambda_k\}\) be the dual variables enforcing \(R_k \ge R_k^{\mathrm{min}}\). The partial Lagrangian can be written as:
\begin{align}
    \mathcal{L}(\{p_{k,j}\}, \{\lambda_k\}) \;=\; 
    \sum_{k=1}^U \sum_{j=1}^U p_{k,j}
    \;+\;\sum_{k=1}^U \lambda_k \bigl(R_k^{\mathrm{min}} - R_k(\{p_{k,j}\})\bigr),
\end{align}
where \(R_k(\{p_{k,j}\})\) is the achievable rate function for user \(k\), derived from the partial MAC (or decoding-order) constraints. By iteratively minimizing \(\mathcal{L}\) w.r.t.\ the powers \(\{p_{k,j}\}\) (a convex problem) and maximizing w.r.t.\ \(\{\lambda_k\}\), one converges to the saddle point that corresponds to the minimal feasible sum power. 

\subsection{Computational Advantages and Practical Use Cases}
The proposed framework decouples power allocation from the combinatorial search over SIC decoding orders, reducing complexity from \((U^2!)^U\) to a manageable set of convex constraints. This enables approximate minimum power estimates without exhaustive order enumeration, supporting real-time scheduling with limited decoding orders or time-sharing among optimal candidates. While not always yielding a single globally optimal order, it provides a practical approach to near-optimal solutions, critical for applications like AR/VR and edge computing that demand rapid feasibility checks and power efficiency.

The approach scales efficiently to large multi-user scenarios, such as 6G networks, where energy efficiency and high data rates are essential. For stringent QoS requirements, it uses dual variables to refine the search, narrowing the permutation space and accelerating convergence.

%% file: results.tex
\section{Results}
\label{sec:results}

\begin{figure}[ht]
    \centering
    \includegraphics[width=0.75\linewidth]{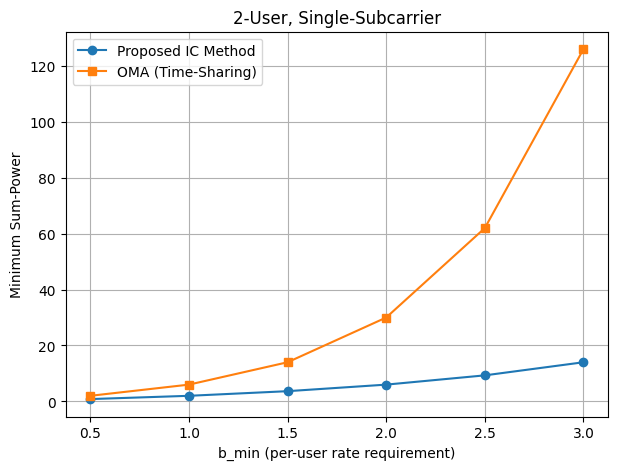}
    \caption{Comparison with OMA}
    \label{fig:results}
\end{figure}

Fig.~\ref{fig:results} shows the minimum sum power obtained for minimum per user data rate requirements, using the multi-user IC optimization, compared to the currently industry standard orthogonal multiple access (OMA) baselines. We obtain higher data rates at a given power level, and conversely, attain minimum required data rates at lower power level compared to industry standard OMA \cite{bhattacharya2025}. We also obtain a 24x lower convergence time using our reformulation.

%% file: conclusion.tex
\section{Conclusion}
\label{sec:conslusion}

In this paper, we developed a framework for characterizing and optimizing the capacity region of multi-user interference channels (IC) using Gaussian inputs. Leveraging the entropy power inequality (EPI), we introduced the partial MAC capacity region to handle selective decoding in ICs. Our construction enumerates finite partial decoding sets without requiring any new auxiliary random variables, thus generalizing (or surpassing) existing bounds such as Han-Kobayashi \cite{HK}, \cite{Sato81} for the multi-user setting.


%% file: appendix.tex
\section{appendix}
\subsection{Introduction}
The Entropy Power Inequality (EPI) is a cornerstone in information theory, instrumental in analyzing communication channel capacities \cite{Verdu2002}. Recent advancements by Wang et. al \cite{Wang_2014} have introduced a refined version of EPI, incorporating spherically symmetric decreasing rearrangements to provide tighter entropy bounds. 

This appendix explores the application of the refined EPI to interference channels with an equal number of users and receivers under additional constraints, including correlated inputs, non-Gaussian noise, and joint entropy considerations across multiple receivers. Mathematical analysis demonstrates how these refinements enhance the understanding of capacity regions and sum-rate bounds in complex multi-user environments. Our goal is to illustrate how the refined EPI enhances bounding techniques, leading to improved insights into the capacity regions and sum-rate bounds of complex multi-user communication systems.




\subsection{Interference Channel Model}

Consider a \( K \)-user interference channel comprising \( K \) transmitters and \( K \) receivers. Each transmitter \( j \) intends to send an independent message to its corresponding receiver \( j \). The channel can be mathematically modeled as:

\begin{equation}
    Y_i = \sum_{j=1}^K h_{ij} X_j + Z_i, \quad \text{for } i = 1, 2, \dots, K
\end{equation}

\noindent where:
\begin{itemize}
    \item \( X_j \) is the transmitted signal from transmitter \( j \),
    \item \( h_{ij} \) is the channel gain from transmitter \( j \) to receiver \( i \),
    \item \( Y_i \) is the received signal at receiver \( i \),
    \item \( Z_i \) is the additive noise at receiver \( i \), typically modeled as Gaussian \( \mathcal{N}(0, N_i) \).
\end{itemize}

\noindent Other \textbf{assumptions} are as follows:
\begin{enumerate}
    \item \textbf{Independent Inputs}: The transmitted signals \( X_j \) are independent across transmitters.
    \item \textbf{Power Constraints}: Each transmitter \( j \) is subject to a power constraint \( \mathbb{E}[X_j^2] \leq P_j \).
    \item \textbf{Gaussian Noise}: The noise terms \( Z_i \) are independent Gaussian random variables.
\end{enumerate}

\subsection{Spherically Symmetric Decreasing Rearrangements}

A \textbf{spherically symmetric decreasing rearrangement} of a function \( f: \mathbb{R}^n \rightarrow \mathbb{R} \) is a rearranged function \( f^* \) that satisfies:

\begin{enumerate}
    \item \textbf{Spherical Symmetry}: 
    \[
    f^*(x) = g(\|x\|) 
    \]
    for some non-increasing function \( g: \mathbb{R} \rightarrow \mathbb{R} \). This means that \( f^* \) depends only on the distance from the origin.
    
    \item \textbf{Non-Increasing Radial Profile}: 
    As the radius \( r = \|x\| \) increases, \( g(r) \) does not increase.
    
    \item \textbf{Preservation of Measure}: 
    For any \( t \geq 0 \),
    \[
    \mu\left( \{ x : f(x) > t \} \right) = \mu\left( \{ x : f^*(x) > t \} \right)
    \]
    where \( \mu \) denotes the Lebesgue measure. This ensures that \( f^* \) is a rearrangement of \( f \) with the same "amount" of mass above any threshold \( t \).
\end{enumerate}
Rearrangements allow us to compare the entropy of complex, potentially asymmetric distributions to their symmetric counterparts. This comparison is crucial in establishing tighter entropy bounds, as symmetric distributions often maximize or minimize entropy under certain constraints.

\subsection{A Refined Entropy Power Inequality (EPI)}

The \textbf{Entropy Power Inequality} provides a lower bound on the entropy of the sum of independent random variables in terms of their individual entropies. Formally, for independent random vectors \( X \) and \( Y \) in \( \mathbb{R}^n \):
\begin{equation}
    N(X + Y) \geq N(X) + N(Y),
\end{equation}
where the \textbf{entropy power} \( N(X) \) is defined as
\begin{equation}
    N(X) = \frac{1}{2\pi e} e^{\frac{2}{n} h(X)},
\end{equation}
and \( h(X) \) is the \textbf{differential entropy} of \( X \).

Madiman and Wang \cite{Wang_2014} introduce a refinement of the classical EPI by incorporating \textbf{spherically symmetric decreasing rearrangements} \cite{Brascamp1974} of random variables. Specifically, for independent random vectors \( X_1, X_2, \ldots, X_n \) in \( \mathbb{R}^d \) with corresponding densities \( f_1, f_2, \ldots, f_n \) and rearrangements \( f_i^* \), the refined EPI states:
\begin{equation}
    h\!\Bigl(\sum_{i=1}^n X_i\Bigr) \;\ge\; h\!\Bigl(\sum_{i=1}^n X_i^*\Bigr),
\end{equation}
where \( X_i^* \) are independent random vectors with spherically symmetric decreasing rearranged densities corresponding to \( X_i \).  
This refinement can provide a tighter lower bound on the entropy of the sum in more intricate multi-user scenarios.

\subsection{Application of Refined EPI to Multi-User Interference Channels}

The refined EPI can be used to derive outer bounds on the sum-rate capacity under various novel conditions: \emph{(i)} correlated inputs, \emph{(ii)} non-Gaussian noise, and \emph{(iii)} joint-entropy considerations across multiple receivers. We illustrate these next.

Application of the refined EPI to interference channels derives outer bounds on the sum-rate capacity under novel conditions: correlated inputs, non-Gaussian noise, and joint-entropy considerations across multiple receivers. The analysis unfolds in three primary cases, each introducing new constraints that deviate from standard assumptions.

\subsubsection{Case 1: Correlated Inputs}

Traditionally, interference channel analyses assume that the transmitted signals \( X_j \) are independent. However, in practical scenarios, transmitters might employ coordinated strategies that lead to partial or full correlation among their signals. Suppose the transmitted signals are jointly Gaussian with a covariance matrix \( \Sigma \):

\[
\bm{X} = (X_1, X_2, \ldots, X_K)^T \sim \mathcal{N}(\bm{0}, \Sigma)
\]

where \( \Sigma \) is a positive semi-definite matrix with at least one non-zero off-diagonal element to indicate correlation between different transmitters.

In a model with correlated inputs, we can first decompose the covariance matrix \( \Sigma \) using the Cholesky decomposition. For simplicity, assume the covariance matrix is full-rank and can be expressed as \( \Sigma = A A^T \), where \( A \) is a lower triangular matrix. Thus, the transmitted signals can be written as
$\bm{X} = A \bm{U}$, where \( \bm{U} = (U_1, U_2, \ldots, U_K)^T \) is a vector of \textit{independent} Gaussian random variables with \( \mathbb{E}[U_i] = 0 \) and \( \mathbb{E}[U_i U_j] = \delta_{ij} P_i \), where \( P_i \) is the power allocated to \( U_i \).


The presence of correlated inputs complicates the direct application of EPI, which typically requires independence. An approach to correlation decomposes the correlated input vector \( \bm{X} \) into a linear combination of independent Gaussian components. Specifically, each transmitted signal \( X_j \) can be written as:
\[
X_j = \sum_{k=1}^K A_{jk} U_k
\]

where \( A_{jk} \) are the elements of matrix \( A \).

The received signal at receiver \( j \) is then:
\begin{align*}
    Y_j &= \sum_{i=1}^K h_{ji} X_i + Z_j \\
        &= \sum_{i=1}^K h_{ji} \left( \sum_{k=1}^K A_{ik} U_k \right) + Z_j \\
        &= \sum_{k=1}^K \left( \sum_{i=1}^K h_{ji} A_{ik} \right) U_k + Z_j
\end{align*}
The effective channel coefficients are defined as:
$c_{jk} = \sum_{i=1}^K h_{ji} A_{ik}$

So, the received signal becomes:
\[
Y_j = \sum_{k=1}^K c_{jk} U_k + Z_j
\]

Since \( \{U_k\} \) are independent Gaussian random variables, the sum \( S_j = \sum_{k=1}^K c_{jk} U_k \) is also Gaussian with variance:

\[
\text{Var}(S_j) = \sum_{k=1}^K c_{jk}^2 P_k
\]

Applying the refined EPI to \( S_j + Z_j \), we obtain:

\[
h(Y_j) = h(S_j + Z_j) \geq h(S_j^* + Z_j^*)
\]

where \( S_j^* \) and \( Z_j^* \) are the spherically symmetric decreasing rearrangements of \( S_j \) and \( Z_j \), respectively. Since \( S_j \) and \( Z_j \) are Gaussian, their rearrangements \( S_j^* \) and \( Z_j^* \) are identical to themselves. Therefore, in this specific case with Gaussian inputs and noise, the refined EPI does not provide a tighter bound than the classical EPI. However, the framework remains valuable when dealing with non-Gaussian or dependent inputs. This is especially useful in applications that communicate different modalities of data like images, text and video, which often have inherent redundancy \cite{compgen25} (or more conceretely, correlation) in the data.

Assuming Gaussian inputs maximize entropy under the given power constraints, \( h(S_j) \) satisfies:
\[
h(S_j) \leq \frac{1}{2} \log\left( 2\pi e \sum_{k=1}^K c_{jk}^2 P_k \right)
\]

Thus, the entropy of the received signal \( Y_j \) satisfies:
\begin{align}
    h(Y_j) &\leq \frac{1}{2} \log\left( 2\pi e \left( \sum_{k=1}^K c_{jk}^2 P_k + N_j \right) \right)
\end{align}

The conditional entropy \( h(Y_j | X_j) \) is similarly bounded. Conditioning on \( X_j \), the received signal reduces to:
\begin{align}
    h(Y_j | X_j) &= h\left( \sum_{i \neq j} h_{ji} X_i + Z_j \right) \\
    &\leq \frac{1}{2} \log\left( 2\pi e \left( \sum_{i \neq j} \sum_{k=1}^K h_{ji}^2 A_{ik}^2 P_k + N_j \right) \right)
\end{align}

The mutual information \( I(X_j; Y_j) \) then satisfies:
\begin{align}
    I(X_j; Y_j) &= h(Y_j) - h(Y_j | X_j) \\
    &\leq \frac{1}{2} \log\left( \frac{ 2\pi e \left( \sum_{k=1}^K c_{jk}^2 P_k + N_j \right) }{ 2\pi e \left( \sum_{i \neq j} \sum_{k=1}^K h_{ji}^2 A_{ik}^2 P_k + N_j \right) } \right) \\
    &= \frac{1}{2} \log\left( \frac{ \sum_{k=1}^K \left( \sum_{i=1}^K h_{ji} A_{ik} \right)^2 P_k + N_j }{ \sum_{i \neq j} \sum_{k=1}^K h_{ji}^2 A_{ik}^2 P_k + N_j } \right) \\
    &= \frac{1}{2} \log\left( 1 + \frac{ \sum_{k=1}^K \left( \sum_{i=1}^K h_{ji} A_{ik} \right)^2 P_k }{ \sum_{i \neq j} \sum_{k=1}^K h_{ji}^2 A_{ik}^2 P_k + N_j } \right)
\end{align}

\noindent Summing over all users, the \textbf{sum-rate} \( R_{\text{sum}} \) is bounded by:

\begin{align}
    R_{\text{sum}} &= \sum_{j=1}^K I(X_j; Y_j) \notag \\
    &\leq \sum_{j=1}^K \frac{1}{2} \log\left( 
        1 + \frac{
            \sum_{k=1}^K \left( \sum_{i=1}^K h_{ji} A_{ik} \right)^2 P_k 
        }{
            \sum_{i \neq j} \sum_{k=1}^K h_{ji}^2 A_{ik}^2 P_k + N_j 
        } 
    \right)
\end{align}

\noindent This bound encapsulates the influence of input correlations through the \( A_{ik} \) coefficients, providing a more refined and accurate upper bound on the sum-rate capacity of the interference channel.
\subsubsection{Case 2: Non-Gaussian Noise}


In many practical scenarios, the noise in communication channels deviates from the ideal Gaussian assumption. Non-Gaussian noise can be impulsive or exhibit heavy-tailed characteristics, such as \(\alpha\)-stable distributions with characteristic exponent \(\alpha < 2\). Let \( Z_i \sim \eta_i \), where \( \eta_i \) represents a non-Gaussian noise distribution.






Assuming that the input signals \( X_j \) remain Gaussian to maximize \( h(S_j) \), \( h(S_j) \) satisfies:
\[
h(S_j) \leq \frac{1}{2} \log\left( 2\pi e \sum_{k=1}^K c_{jk}^2 P_k \right)
\]

For non-Gaussian noise \( Z_j \), the entropy \( h(Z_j^*) \) reflects the rearranged distribution's characteristics. The refined EPI provides a lower bound on \( h(Y_j) \):
\[
h(Y_j) \leq \frac{1}{2} \log\left( 2\pi e \left( \sum_{k=1}^K c_{jk}^2 P_k + N_j^* \right) \right)
\]

where \( N_j^{\ast} \) is the entropy power of the rearranged non-Gaussian noise \( Z_j^* \).

The mutual information \( I(X_j; Y_j) \) is then bounded by:
\begin{align}
    I(X_j; Y_j) &= h(Y_j) - h(Y_j | X_j) \\
    &\leq \frac{1}{2} \log\left( \frac{ 2\pi e \left( \sum_{k=1}^K c_{jk}^2 P_k + N_j^* \right) }{ 2\pi e \left( \sum_{i \neq j} \sum_{k=1}^K h_{ji}^2 A_{ik}^2 P_k + N_j \right) } \right) \\
    &= \frac{1}{2} \log\left( \frac{ \sum_{k=1}^K \left( \sum_{i=1}^K h_{ji} A_{ik} \right)^2 P_k + N_j^* }{ \sum_{i \neq j} \sum_{k=1}^K h_{ji}^2 A_{ik}^2 P_k + N_j } \right) \\
    &= \frac{1}{2} \log\left( 1 + \frac{ \sum_{k=1}^K \left( \sum_{i=1}^K h_{ji} A_{ik} \right)^2 P_k }{ \sum_{i \neq j} \sum_{k=1}^K h_{ji}^2 A_{ik}^2 P_k + N_j } \right)
\end{align}

\noindent Summing over all users, the \textbf{sum-rate} \( R_{\text{sum}} \) is bounded by:
\begin{align}
    R_{\text{sum}} &= \sum_{j=1}^K I(X_j; Y_j) \notag \\
    &\leq \sum_{j=1}^K \frac{1}{2} \log\left( 
        1 + \frac{
            \sum_{k=1}^K 
            \left( 
                \sum_{i=1}^K h_{ji} A_{ik} 
            \right)^2 P_k
        }{
            \sum_{i \neq j} 
            \sum_{k=1}^K h_{ji}^2 A_{ik}^2 P_k + N_j 
        }
    \right)
\end{align}


\subsubsection{Case 3: Joint Entropy Bounds Across Receivers}

While individual entropy bounds provide insights into each receiver's capacity, considering the joint entropy of multiple receivers can yield \textbf{tighter} and more comprehensive capacity region bounds. This is particularly relevant in scenarios involving network MIMO or cooperative decoding architectures, where joint processing of received signals is feasible.


For a vector of received signals \( \mathbf{Y} = (Y_1, Y_2, \ldots, Y_K) \), the joint entropy \( h(\mathbf{Y}) \) can be bounded using the refined EPI in a multi-dimensional setting. The refined EPI can be extended to handle sums of vector-valued random variables, provided the independence conditions are appropriately generalized.

Applying the refined EPI to \( \mathbf{Y} \), we obtain:

\[
h(\mathbf{Y}) = h\left( \sum_{j=1}^K \mathbf{h}_j X_j + \mathbf{Z} \right) \geq h\left( \sum_{j=1}^K \mathbf{h}_j^* X_j^* + \mathbf{Z}^* \right)
\]

\noindent where \( \mathbf{h}_j \) represents the channel gains vector for transmitter \( j \), and \( \mathbf{Z} \) is the noise vector across all receivers.


Assuming Gaussian inputs and leveraging the refined EPI, the joint entropy \( h(\mathbf{Y}) \) satisfies:

\[
h(\mathbf{Y}) \leq \frac{1}{2} \log\left( (2\pi e)^K \det\left( \sum_{j=1}^K h_{jj}^2 P_j \mathbf{I} + \mathbf{N} \right) \right)
\]

where \( \mathbf{I} \) is the identity matrix and \( \mathbf{N} \) is the noise covariance matrix.

The joint mutual information across all receivers can then be bounded as:

\begin{align}
    I(\mathbf{X}; \mathbf{Y}) &= h(\mathbf{Y}) - h(\mathbf{Y} | \mathbf{X}) \\
    &\leq \frac{1}{2} \log\left( \det\left( \sum_{j=1}^K h_{jj}^2 P_j \mathbf{I} + \mathbf{N} \right) \right) - \frac{1}{2} \log\left( \det\left( \mathbf{N} \right) \right) \\
    &= \frac{1}{2} \log\left( \det\left( \mathbf{I} + \mathbf{N}^{-1} \sum_{j=1}^K h_{jj}^2 P_j \mathbf{I} \right) \right)
\end{align}

Summing over all users, the \textbf{sum-rate} \( R_{\text{sum}} \) is bounded by:

\begin{equation}
    R_{\text{sum}} \leq \frac{1}{2} \log\left( \det\left( \mathbf{I} + \mathbf{N}^{-1} \sum_{j=1}^K h_{jj}^2 P_j \mathbf{I} \right) \right)
\end{equation}

\noindent This joint entropy bound captures the collective behavior of the interference channel, potentially leading to tighter capacity region outer bounds compared to treating each receiver independently. It accounts for the interdependencies among received signals, which is particularly beneficial in scenarios where joint decoding is feasible.
\subsection{Further Work}


Building on the above appendix, ongoing work involves applying these refined bounds in higher-dimensional settings, exploring advanced coding strategies to match or approach these refined outer bounds, and validating performance gains through numerical simulation in real-world system models.

